\documentclass[aps,prb,preprint,superscriptaddress,floatfix,showpacs]{revtex4}
\usepackage{graphicx}
\usepackage{dcolumn}
\usepackage{amsmath}
\usepackage{color}

\begin{document}

\title{Computation of correlation-induced atomic displacements and 
structural transformations in paramagnetic KCuF$_3$ and LaMnO$_3$}

\author{I. Leonov}
\affiliation{Theoretical Physics III, Center for Electronic Correlations and Magnetism,
Institute of Physics, University of Augsburg, Augsburg 86135, Germany}
\author{Dm. Korotin}
\affiliation{Institute of Metal Physics, S. Kovalevskoy St. 18, 620219 Yekaterinburg
GSP-170, Russia}
\author{N. Binggeli}
\affiliation{Abdus Salam International Center for Theoretical Physics
and INFM-CNR Democritos National Simulation Center, Trieste 34014, Italy}
\author{V. I. Anisimov}
\affiliation{Institute of Metal Physics, S. Kovalevskoy St. 18, 620219 Yekaterinburg
GSP-170, Russia}
\author{D. Vollhardt}
\affiliation{Theoretical Physics III, Center for Electronic Correlations and Magnetism,
Institute of Physics, University of Augsburg, Augsburg 86135, Germany}

\begin{abstract}
We present a computational scheme for \textit{ab initio}
total-energy calculations of materials with strongly interacting
electrons using a plane-wave basis set. It combines \textit{ab initio} band
structure and dynamical mean-field theory and is implemented in terms of
plane-wave pseudopotentials. The present approach allows us to investigate
complex materials with strongly interacting electrons and is able to treat
atomic displacements, and hence structural transformations, caused by
electronic correlations. Here it is employed to investigate two prototypical 
Jahn-Teller materials,
KCuF$_3$ and LaMnO$_3$, in their paramagnetic phases. The computed equilibrium
Jahn-Teller distortion and antiferro-orbital order agree well with experiment, 
and the structural
optimization performed for paramagnetic KCuF$_3$ yields the correct
lattice constant, equilibrium Jahn-Teller distortion and tetragonal compression
of the unit cell. Most importantly, the present approach is able to determine
correlation-induced structural transformations, equilibrium atomic
positions and lattice structure in both strongly and weakly correlated
solids in their \textit{paramagnetic} phases as well as in phases with
long-range magnetic order.
\end{abstract}

\maketitle

\section*{I.  INTRODUCTION}

The theoretical understanding of complex materials with strongly interacting electrons
is one of the most challenging areas of current research in condensed matter physics.
Experimental studies of such materials have often revealed rich phase diagrams
originating from the interplay between electronic and lattice degrees of freedom. \cite{Rev}
This makes these compounds particularly interesting in view of possible technological 
applications.
Namely, the great sensitivity of many correlated electron materials with respect to
changes of external parameters such as temperature, pressure, magnetic and/or electric
field, doping, etc., can be employed to construct materials with useful 
functionalities. \cite{Rev}

The electronic properties of materials can be computed from first principles by density
functional theory in the local density approximation (LDA), \cite{LDA} the generalized
gradient approximation (GGA), \cite{PB96,GGALDA} or using the so-called LDA+U
method. \cite{AZ91,LA95}
Applications of these approaches accurately describe
the phase diagrams of many simple elements and semiconductors, and of some insulators.
Moreover, they often allow to make correct qualitative predictions of the magnetic,
orbital, and crystal structures of solids, where the equilibrium (thermodynamic) structures 
are determined by simultaneous optimization of the 
electron and lattice systems. \cite{RevFLAPW,RevPWSCF,RevWIEN2k,RevVASP,RevABINIT,RevSIESTA}
However, these methods usually fail to describe the correct electronic and structural
properties of electronically correlated \textit{paramagnetic} materials.
Hence the computation of electronic, magnetic, and structural properties of strongly 
correlated paramagnetic materials remains a great challenge. 

Here the recently
developed combination of conventional band structure theory and dynamical mean-field
theory, \cite{DMFT} the so-called LDA+DMFT computational scheme, \cite{DMFTmeth}
has become a powerful tool for the investigation of strongly
correlated compounds in both their paramagnetic and magnetically ordered states. 
In particular, it provides important
insights into the spectral and magnetic properties of correlated electron
materials, \cite{DMFTcalc,DMFTcalc+} especially in the vicinity of a Mott
metal-insulator transition as encountered in transition metal oxides. \cite{Rev}
Up to now implementations of the LDA+DMFT approach utilized
linearized and higher-order muffin-tin orbital [L(N)MTO] techniques \cite{LNMTO}
and focused on the investigation of electronic correlation effects for a
given lattice structure. However, the mutual interaction between
electrons and ions, i.e., the influence of the electrons on the lattice
structure, is then completely neglected. LDA+DMFT computations of the volume
collapse in paramagnetic Ce \cite{HM01,AB06} and Pu \cite{SKA01+} and of the
collapse of the magnetic moment in MnO \cite{MnO} did include the lattice,
but only calculated the total energy of the correlated material as a
function of the unit cell volume. \cite{DM09+KS09}
In the case of more subtle structural transformations, e.g.,
involving the cooperative Jahn-Teller (JT) effect, \cite{JT37,KK} the L(N)MTO 
technique is not suitable since it cannot reliably determine atomic positions.
This is
due to the atomic-sphere approximation within the L(N)MTO scheme, where a
spherical potential inside the atomic sphere is employed. Thereby multipole
contributions to the electrostatic energy due to the distorted charge density
distribution around the atom are ignored. 
Instead, the recently proposed implementation of the LDA+DMFT approach, which
employs plane-wave pseudopotentials \cite{LB08,TL08,DK08,AL08,LG06} and thus 
avoids the atomic sphere approximation, does not neglect such contributions. 
Thereby it becomes possible to describe the effect of the distortion
on the electrostatic energy. \cite{LB08}

In this paper, we present a detailed formulation of the LDA+DMFT scheme 
implemented with 
plane-wave pseudopotentials reported earlier. \cite{LB08,TL08} This scheme 
allows us to compute structural transformations (e.g., structural 
phase stability and structure optimization) caused by electronic correlations. 
Most importantly, it is able to
determine correlation-induced structural transformations in both 
paramagnetic solids and long-range ordered solids.
Therefore, the present computational scheme overcomes the 
limitations of standard band-structure approaches and opens the way for 
fully microscopic
investigations of the structural properties of strongly correlated electron
materials. 

We apply this method to 
study orbital order and the cooperative JT distortion in two prototypical
JT materials, KCuF$_3$ and LaMnO$_3$, and compute the electronic, structural, 
and orbital properties in their room-temperature paramagnetic phase.
At room temperature, both compounds have a distorted perovskite structure with a
strong cooperative JT distortion. Considering this structural complexity, KCuF$_3$
has a relatively high (tetragonal) symmetry, \cite{BM90} where only the
cooperative JT distortion of the CuF$_6$ octahedra is responsible for the
deviation from the cubic perovskite symmetry. LaMnO$_3$ instead crystallizes
in a more complex (orthorhombic) structure in which MnO$_6$ octahedra are
simultaneously JT distorted and tilted with respect to the ideal cubic
perovskite structure. \cite{EL71,RH98,CF03} The JT distortion persists up to
the melting temperature $\sim 1000$ K in KCuF$_3$. By contrast, in LaMnO$_3$
it persists only up to $T_{JT} \sim 750$ K, the temperature at which the 
JT distortion vanishes and where
LaMnO$_3$ undergoes a structural phase transition with a volume collapse 
to a nearly cubic structure without JT distortion and orbital 
order. \cite{RH98,CF03} Concerning the electronic configuration, KCuF$_3$ 
nominally has a Cu$^{2+}$ $3d^9$ structure, i.e., a single hole
in the $3d$ shell. By contrast, due to Hund's rule coupling
LaMnO$_3$ has a single $e_{g\uparrow}$ electron
bound to the fully spin-polarized three $t_{2g\uparrow}$ electrons in 
a high spin $3d^4$ ($t_{2g\uparrow}^3e_{g\uparrow}^1$) electronic configuration. 
To properly describe this correlated state,
a different treatment compared to KCuF$_3$ is required, which takes 
into account the effective on-site spin interaction between $t_{2g}$ and $e_g$ 
electrons. \cite{MS96,HV00} At low temperatures,
both systems display $A$-type long-range antiferromagnetic order,
consistent with the Goodenough-Kanamori-Anderson rules for a superexchange
interaction with antiferro-orbital order. In both compounds the 
N\' eel temperature 
($T_N\sim 38$ K in KCuF$_3$ and $T_N\sim 140$ K in LaMnO$_3$) is remarkably lower
than $T_{JT}$.

In this paper we will show that our approach can explain the orbital order, 
cooperative JT distortion, and related structural properties in both materials, 
in spite of their chemical, structural, and electronic differences. The scheme 
is robust and makes it possible to address, on the same footing, electronic, 
magnetic, and structural properties of strongly correlated materials.

The paper is organized as follows. In Section II we present computational
details needed to reproduce the results of our calculations. The crystal
structures, magnetic properties, and results of electronic structure
calculations of paramagnetic KCuF$_3$ and LaMnO$_3$ are presented in
Sections III and IV, respectively. Finally, the results are summarized in
Section V.


\section*{II.  COMPUTATIONAL DETAILS}

In order to compute the electronic structure of correlated electron materials,
we have implemented DMFT within a realistic electronic structure approach,
which is formulated in terms of plane-wave pseudopotentials. \cite{LB08,TL08}
Following the
paper of Anisimov \emph{et al.} \cite{AK05} and Trimarchi \emph{et al.}, \cite{TL08} 
this can be achieved by applying a projection onto atomic-centered
symmetry-constrained Wannier orbitals, \cite{MV97} which gives an effective
low-energy Hamiltonian $\hat H_{DFT}$ for the partially filled correlated
orbitals (e.g., $3d$ orbitals of transition metal ion).

The Hamiltonian $\hat H_{DFT}$ provides a realistic description of the 
material's single-electron band structure.
It is supplemented by on-site Coulomb interactions for
the correlated orbitals, resulting in a many-body Hamiltonian of the
form:
\begin{eqnarray}
{\hat H} & = & {\hat H_{DFT}} + U\sum_{im} \hat n_{im\uparrow} \hat
n_{im\downarrow}  \notag \\
& + & \sum_{i m\neq m^{\prime} \sigma \sigma^{\prime}} (V - \delta_{\sigma
\sigma^{\prime}} J) \hat n_{i m\sigma} \hat n_{i m^{\prime} \sigma^{\prime}}
- {\hat H_{DC}}.
\end{eqnarray}
Here the second and third terms on the right-hand side describe the local
Coulomb interaction between electrons in the same and in different
correlated orbitals, respectively, with $V=U-2J$, and ${\hat H_{DC}}$ is a
double counting correction which accounts for the electronic interactions
already described by DFT (see below). The Coulomb repulsion $U$ and Hund's rule
exchange $J$ can be evaluated using a constrained DFT scheme within
a Wannier-functions formalism, making the Hamiltonian (1) free of 
adjustable parameters. \cite{DK08}

The many-body Hamiltonian (1) is then solved by dynamical mean-field theory
(DMFT) \cite{DMFT} with the effective impurity model treated, e.g., by the
numerically exact Hirsch-Fye quantum Monte-Carlo method. \cite{HF86}
Finally, applying a maximum entropy treatment of Monte-Carlo data, one
obtains the real-frequency spectral functions, which can be further compared
to physically observable spectra.

The total energy is another important quantity which can be evaluated from
DFT+DMFT calculation using the following expression \cite{AB06,LB08}
\begin{equation}
E = E_{DFT}[\rho] + \langle \hat H_{DFT} \rangle - \sum_{m,k}
\epsilon^{DFT}_{m,k} + \langle \hat H_{U} \rangle - E_{DC}.
\label{eqn:energy}
\end{equation}
Here $E_{DFT}[\rho]$ is the total energy obtained by DFT. The third term on
the right-hand side of Eq.~(2) is the sum of the DFT 
valence-state eigenvalues which is evaluated as the thermal average of the
DFT Hamiltonian with the non-interacting DFT Green's function 
$G^{DFT}_{\mathbf{k}}(i\omega_n)$:
\begin{equation}
\sum_{m,k} \epsilon^{DFT}_{m,k} = T\sum_{i\omega_n,\mathbf{k}} 
Tr[H_{DFT}(\mathbf{k}) G^{DFT}_{\mathbf{k}}(i\omega_n)] e^{i\omega_n0^{+}}.
\end{equation}
In this expression, we have assumed that the DFT total energy has only a weak
temperature dependence via the Fermi distribution function, i.e., one
neglects the temperature dependence of the exchange-correlation potential. 
$\langle \hat H_{DFT} \rangle$ is evaluated similarly but with the full
Green's function including the self-energy. To calculate these two
contributions, the summation is performed over the Matsubara frequencies 
$i\omega_n$ (typically with $n_{max}=10^3$), taking into account an
analytically evaluated asymptotic correction (see below). Thus, for 
$\langle \hat H_{DFT} \rangle$ one has
\begin{eqnarray}
&&\langle \hat H_{DFT} \rangle = T\sum_{i\omega_n,\mathbf{k}} Tr[H_{DFT}(
\mathbf{k}) G_{\mathbf{k}}(i\omega_n)] e^{i\omega_n0^{+}}  \notag \\
&=& T\sum^{\pm i\omega_{n_{max}}}_{i\omega_n,\mathbf{k}} Tr \{ H_{DFT}(
\mathbf{k}) [ G_{\mathbf{k}}(i\omega_n)- \frac{m^{\mathbf{k}}_{1}}{%
(i\omega_n)^2} ] \}  \notag \\
&+&\frac{1}{2} \sum_{\mathbf{k}} Tr[H_{DFT}(\mathbf{k})]- \frac{1}{4T} 
\sum_{\mathbf{k}} Tr[H_{DFT}(\mathbf{k}) m^{\mathbf{k}}_{1}]
\end{eqnarray}
where the first moment $m^{\mathbf{k}}_{1}$ is computed as
$m^{\mathbf{k}}_{1} = H_{DFT}(\mathbf{k}) + \Sigma(i\infty)- \mu$;
the asymptotic part of the self-energy $\Sigma(i \infty)$ is calculated as
the average of $\Sigma(i\omega_n)$ over the last several $i\omega_n$ points.
The interaction energy $\langle \hat H_{U} \rangle$ is computed from the
double occupancy matrix. The double-counting correction $E_{DC}$ is
evaluated as the average Coulomb repulsion between the $N_{d}$
correlated electrons in the Wannier orbitals. In the case of a paramagnet
it corresponds to $E_{DC}= \frac{1}{2}U
N_{d}(N_{d}- 1)-\frac{1}{4}J N_{d}(N_{d}- 2)$. Since the Hamiltonian
involves only correlated orbitals the number of Wannier electrons 
$N_{d}$ is conserved. Therefore, the double-counting correction amounts to an
irrelevant shift of the total energy.

Within this approach, we can determine correlation-induced
structural transformations, as well as the corresponding change of the atomic
coordinates and of the unit cell shape. The result can be further used to
explain the experimentally observed structural data and to predict
structural properties of real correlated materials. In 
Sections III and IV we will apply this method to determine the orbital order and
the cooperative JT distortion in the paramagnetic phase of two prototypical JT
systems KCuF$_3$ and LaMnO$_3$.

\section*{III.  APPLICATION TO KCuF$_3$}

\subsection*{A.  Crystal structure and magnetic properties}

KCuF$_3$ is the prototype of a material with a cooperative Jahn-Teller (JT) distortion,
orbital order, and low-dimensional magnetism. \cite{KY67,KK} At room
temperature it crystallizes in a pseudocubic perovskite crystal structure 
\cite{BM90} which is related to the crystal structure of high-$T_c$
superconductors and colossal magnetoresistance manganites and, particularly,
to their parent compound, LaMnO$_3$. 
Due to the particular orbital order in KCuF$_3$ it is one of the rare examples 
of an ideal one-dimensional antiferromagnetic Heisenberg system. \cite{KK} 
Thus, the copper
ions have an octahedral fluorine surrounding and are nominally in a Cu$^{2+}$ 
($3d^9$) electronic configuration with a single hole in the $e_g$ states. This
gives rise to a strong JT instability that lifts the cubic degeneracy at Cu 
$e_g$ states due to a cooperative JT distortion. \cite{KK} The
latter is characterized by CuF$_6$ octahedra elongated along the $a$
and $b$ axis and arranged in an antiferro-distortive pattern in the $ab$
plane. \cite{BM90} At room temperature, there are two different structural
polytypes with antiferro ($a$-type) and ferrolike ($d$-type) stacking of the
$ab$ planes along the $c$ axis. \cite{O69} The JT distortion is associated
with the particular orbital order in KCuF$_3$, in which a single hole alternatingly
occupies $d_{x^2-z^2}$ and $d_{y^2-z^2}$ orbital states, resulting in a
tetragonal compression ($c<a$) of the unit cell. The mechanism responsible for
the orbital order in KCuF$_3$ is still being debated in the
literature. \cite{LA95,KK,G63,MK02,LB08,PK08,BA04}
In particular, purely electronic effects such as in the Kugel-Khomskii theory \cite{KK} 
and the
electron-lattice interaction \cite{G63} have been discussed as possible
driving forces behind the orbital order.

The relatively high (tetragonal) symmetry makes KCuF$_3$ one of the simplest
system to study. In particular, in order to describe the JT distortion, only
a single internal structure parameter, the shift of the in-plane fluorine
atom from the Cu-Cu bond center, is needed. Moreover, KCuF$_3$ has a single
hole in the $3d$ shell resulting in absence of multiplet effects.
Altogether, the electronic and structural properties of KCuF$_3$ have been
intensively studied by density functional theory in the local
density approximation (LDA), \cite{LDA} the generalized gradient approximation
(GGA), \cite{PB96,GGALDA} or using the so-called LDA+U approach. \cite{AZ91,LA95} 
While the LDA+U calculations account rather well for the value
of equilibrium JT distortion in KCuF$_3$, \cite{BA04} the calculations 
simultaneously
predict long-range antiferromagnetic order which indeed occurs in 
KCuF$_3$ below $T_{N}$ ($\sim 38$ K and 22 K for the $a$ polytype and for 
the $d$ polytype, respectively). \cite{HS69} Note, however, that the N\' eel
temperature is much lower than the critical temperature for orbital order
which is generally assumed to be as large as the melting temperature ($\sim 1000$ K).
The LDA+U calculations give the correct insulating ground state with the
long-range $A$-type antiferromagnetic and $d_{x^2-z^2}/d_{y^2-z^2}$
antiferro-orbital order, \cite{LA95,MK02,BA04} consistent with the
Goodenough-Kanamori-Anderson rules for a superexchange interaction.
Non-magnetic LDA calculations instead predict a \emph{metallic} behavior.
Moreover, the electronic and structural properties of KCuF$_3$ have been
recently reexamined by means of LDA+U molecular-dynamic simulations,
indicating a possible symmetry change and challenging the original
assignment of tetragonal symmetry. \cite{BA04} This symmetry change seems to
allow for a better understanding of Raman, \cite{U91} electronic paramagnetic
resonance, \cite{Y89,EZ08} and x-ray resonant scattering \cite{PC02}
properties at $T\approx T_N$. However, the details of this distortion have
not been fully resolved yet.

The LDA+U approach is able to determine electronic properties and
the JT distortion in KCuF$_3$ rather well, \cite{BA04} but the application of
this approach is limited to temperatures below $T_{N}$. LDA+U cannot explain
the properties at $T>T_{N}$ and, in particular, at room temperature, where
KCuF$_3$ is a correlated paramagnetic insulator with a robust JT distortion
which persists up to the melting temperature.

Here we present an application of the GGA+DMFT computational scheme
formulated in terms of plane-wave pseudopotentials \cite{LB08,TL08} which 
allows us to determine the structural properties, in particular, the amplitude 
of the equilibrium JT distortion and its energetics, in \emph{paramagnetic}
KCuF$_3$. We also report results of a structural optimization, at constant
volume (constant external pressure) and lattice symmetry, including 
optimization of the unit cell shape and fluorine atomic positions.

\subsection*{B.  Electronic structure and orbital order}


In this section, we present results of the GGA and GGA+DMFT electronic
structure calculations of paramagnetic insulator KCuF$_3$. In these
calculations we have used the experimental room-temperature crystal
structure with space group $I4/mcm$ and lattice constants $a=5.855$ 
and $c=7.852$ \AA . \cite{BM90} The calculations were performed for 
different values of the JT distortion defined accordingly as 
$\delta_{\mathrm{JT}}= \frac{1}{2}(d_l-d_s)/(d_l+d_s)$ where
$d_l$ and $d_s$ denote the long and short Cu-F bond distances in the $ab$
plane of CuF$_6$ octahedra, respectively, and 
$2(d_l+d_s)=a$. In the following we express the distortion $\delta_{\mathrm{JT}}$
in percent of the lattice constant $a$, e.g., $\delta_{\mathrm{JT}}=0.002 \equiv 0.2$ \%.
In our investigation  we consider $0.2 \% \leq \delta_{\mathrm{JT}} \leq 7 \%$. 
The structural
data \cite{BM90} at room-temperature yield $\delta_{\mathrm{JT}}= 4.4$ \%. 
In the present
calculations we keep the lattice parameters $a$ and $c$ and the space group
symmetry fixed, whereas the structural optimization involving change of both
the JT distortion and lattice constants will be discussed in the following
section.

We first calculate the non-magnetic GGA electronic structure of KCuF$_3$,
employing the plane-wave pseudopotential approach. \cite{PSEUDO,PB96} For
these calculations we use the Perdew-Burke-Ernzerhof exchange-correlation
functional together with Vanderbilt ultrasoft pseudopotentials for copper 
and fluorine, and a soft Troullier-Martin pseudopotential for potassium.
The nonlinear core correction to the exchange-correlation potential has been
included to account for the overlap between the valence and core electrons.
All calculations are carried out in a tetragonal unit cell which contains two
formula units (10 atoms) per primitive unit cell. We use a kinetic-energy
cutoff of 75 Ry for the plane-wave expansion of the electronic states. The
integration in reciprocal space is performed using a [8,8,8] Monkhorst-Pack 
$k$-point grid.

\begin{figure}[tbp!]

\centerline{\includegraphics[width=0.45\textwidth,clip]{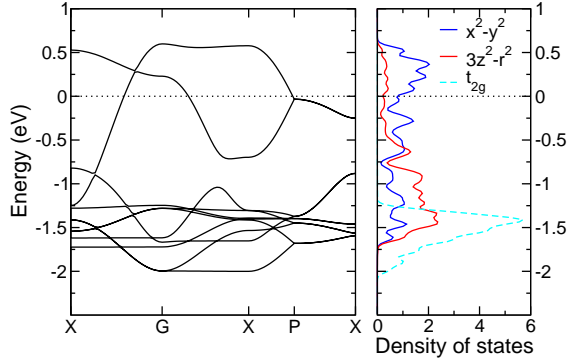}}
\caption{(color online) Band structure and orbitally resolved 
Cu $t_{2g}$ and $e_g$ spectral densities of paramagnetic KCuF$_3$ 
as obtained by non-magnetic GGA for $\protect\delta_{\mathrm{JT}}= 4.4 \%$. 
The zero of energy corresponds to the Fermi level.}
\label{fig:kcuf_ggados}
\end{figure}

For all values of $\delta_{\mathrm{JT}}$ considered here, the non-magnetic 
GGA yields a \textit{metallic} rather than the experimentally observed 
insulating
behavior, with an appreciable orbital polarization due to the crystal field
splitting. This is shown in Fig.~\ref{fig:kcuf_ggados} which presents the
GGA density of states and the corresponding band structure calculated for 
$\delta_{\mathrm{JT}}=4.4$ \%. Overall, the non-magnetic GGA results qualitatively
agree with previous band-structure calculations. \cite{LA95,MK02,BA04} The
Cu $t_{2g}$ states are completely occupied and located at about 1-2 eV below
the Fermi level. Partially filled bands at the Fermi level originate from
the Cu $e_g$ orbitals. We note that an increase of the JT distortion results in
a considerable enhancement of the crystal field splitting between $x^2-y^2$ 
and $3z^2- r^2$ bands (in the local frame \cite{dif_orb}) that leads to 
an appreciable depopulation of the $x^2-y^2$ orbital. There is a minor
narrowing of the $t_{2g}$ and $e_g$ bands of $\sim 0.2$ and 0.1 eV,
respectively, as well as a slight up-shift of the center of gravity of the 
$t_{2g}$ bands ($\sim 0.1$ eV) with decreasing JT distortion.

\begin{figure}[tbp!]
\centerline{\includegraphics[width=0.45\textwidth,clip]{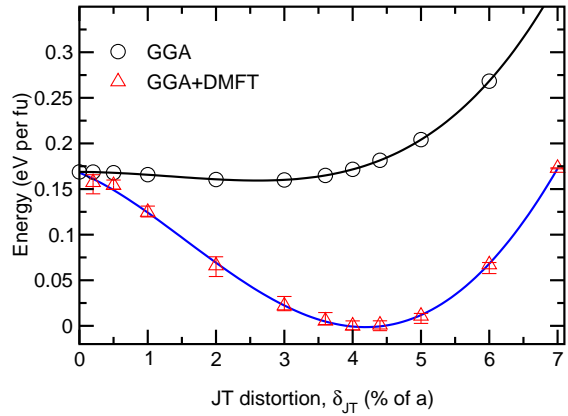}}
\caption{(color online) Comparison of the total energies of paramagnetic 
KCuF$_3$ computed by GGA and GGA+DMFT(QMC) as a function of the JT distortion.
Error bars indicate the statistical error of the DMFT(QMC) calculations. }
\label{fig:kcuf_energy}
\end{figure}

In Fig.~\ref{fig:kcuf_energy} we display our results for the GGA total
energy as a function of the JT distortion $\delta_{\mathrm{JT}}$. 
Notice that, in
agreement with previous studies, \cite{LA95,BA04} the electron-lattice
interaction alone is found insufficient to stabilize the orbitally ordered
insulating state. The non-magnetic GGA calculations not only give a metallic
solution, but its total energy profile is seen to be almost constant for 
$\delta_{\mathrm{JT}}< 4$ \%, with a very shallow minimum at about 2.5 \%. 
This would
imply that KCuF$_3$ has no JT distortion for temperature above 100 K, which
is in clear contradiction to experiment. Obviously, a JT distortion by itself,
without the inclusion of electronic correlations, cannot explain the
experimentally observed orbitally ordered \emph{insulating} state in
paramagnetic KCuF$_3$.

\begin{figure}[tbp!]
\includegraphics[width=0.22\textwidth]{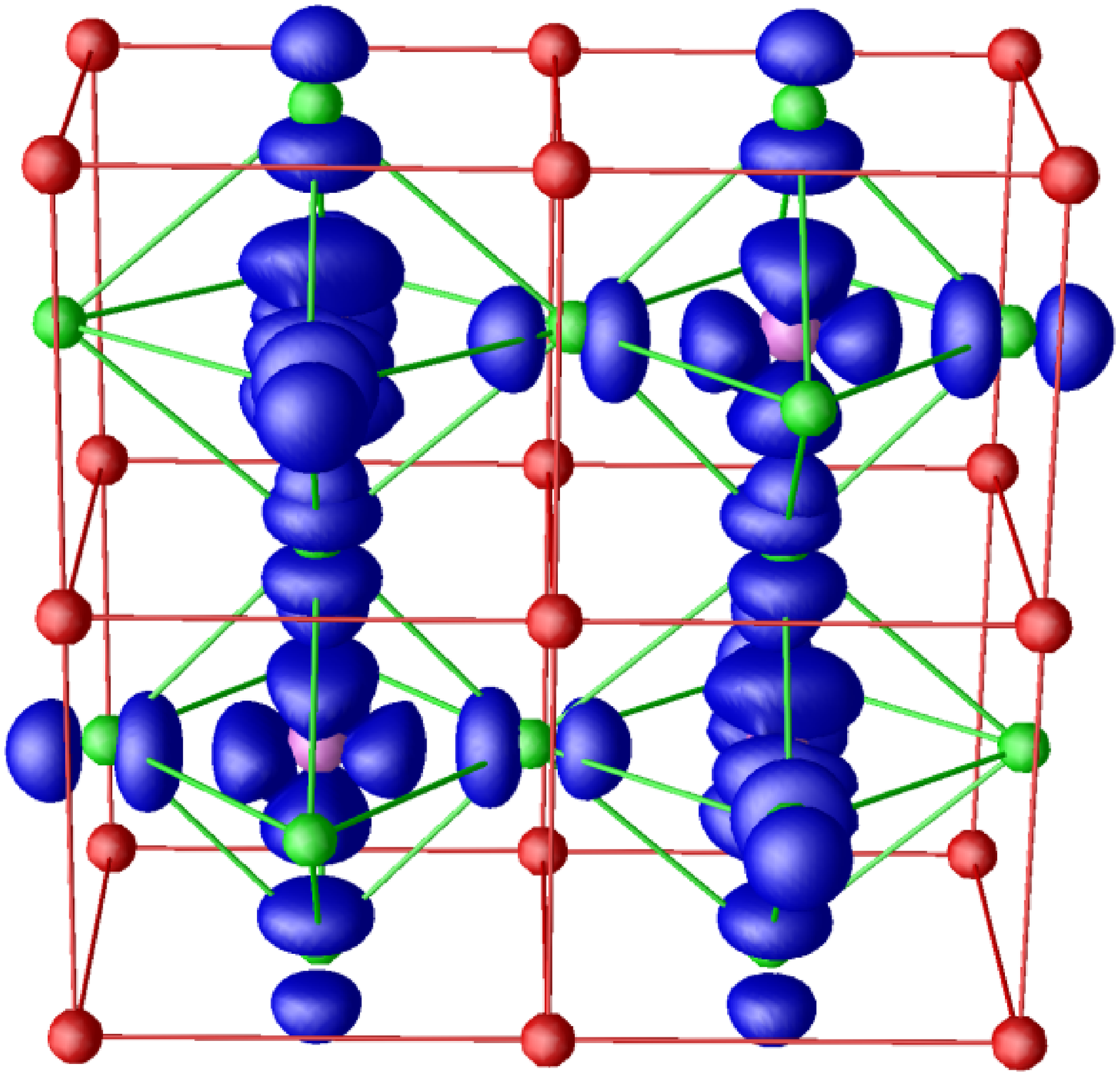}
\includegraphics[width=0.23\textwidth]{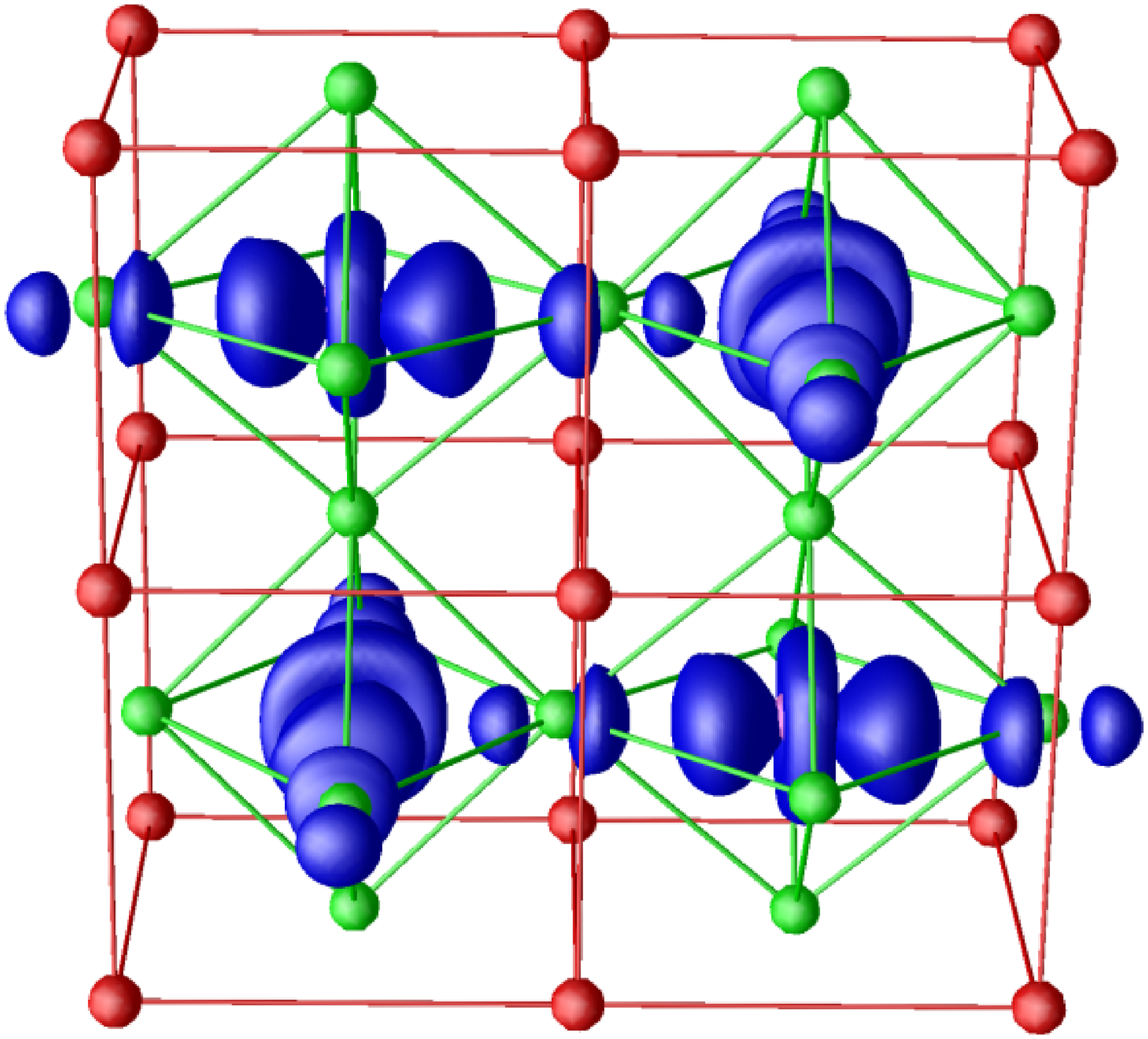}
\caption{(color online) $I4/mcm$ primitive cell and Wannier $e_g$
orbitals ($x^2-y^2$ and $3z^2-r^2$ in left and right, respectively) as
obtained by non-magnetic GGA for KCuF$_3$ with 
$\protect\delta_{\mathrm{JT}}=4.4$ \%. The fluorine atoms and fluorine octahedra 
are shown in green, the potassium in red, and the Wannier Cu $e_g$ charge density 
in blue. The local coordinate system is chosen with the \emph{z} direction defined 
along the longest Cu-F bond of the CuF$_6$ octahedron. }
\label{fig:kcuf_wan}
\end{figure}

To proceed further, we consider the partially filled Cu $e_g$ orbitals as
correlated orbitals and construct an effective low-energy Hamiltonian ${\hat
H_{DFT}}$ for each value of the JT distortion $\delta_{\mathrm{JT}}$ considered
above. This is achieved by employing the pseudopotential plane-wave GGA
results and making a projection onto atomic-centered symmetry-constrained Cu
$e_g$ Wannier orbitals. \cite{TL08} The resulting Cu $x^2-y^2$ and $3z^2-r^2$
Wannier orbitals calculated for $\delta_{\mathrm{JT}}=4.4$ \% are shown 
Fig.~\ref{fig:kcuf_wan}. Taking the local Coulomb repulsion $U=7$ eV and 
Hund's rule
exchange $J=0.9$ eV \cite{LA95} into account, we obtain the many-body
low-energy Hamiltonian (1) for the two $(m=1,2)$ Cu $e_g$ orbitals, which is
further solved (for each value of $\delta_{\mathrm{JT}}$) within the single-site 
DMFT using Hirsch-Fey quantum Monte Carlo (QMC) calculations. 
\cite{HF86,Off-diag-elements} The calculations have been performed at $T=1160$ K
($\beta=10$ eV$^{-1}$), using 40 imaginary-time slices. The Matsubara sums in
Eq. 2 have been taken over $n_{max}=10^3$ frequencies; this gives accuracy
in the total energy calculation better than 10 meV per formula unit.

Using now the expression of Eq.~\ref{eqn:energy}, we have calculated the
GGA+DMFT total energy for each value of the JT distortion 
$\delta_{\mathrm{JT}}$
considered here. The result of the paramagnetic GGA+DMFT computation of the
total energy is presented in Fig.~\ref{fig:kcuf_energy}, where it is
compared with the non-magnetic GGA calculation. In contrast to the GGA
result, the inclusion of the electronic correlations among the partially
filled Cu $e_g$ states in the GGA+DMFT approach leads to a very substantial
lowering of the total energy by $\sim$ 175 meV per formula unit. This
implies that the strong JT distortion persists up to the melting temperature
($>1000$ K), in agreement with experiment. This finding is in strong
contrast to the absence of JT distortion above 100 K predicted by
GGA. The minimum of the GGA+DMFT total energy is located at the value 
$\delta_{\mathrm{JT}}\approx 4.2$ \%, which is also in excellent agreement 
with the
experimental value of 4.4 \%. \cite{BM90} Note however that the total energy
minimum position depends on the value of Coulomb interaction parameter U.
Thus, the calculations of the total energy minima for $U=6$ eV and $U=8$
eV result in optimal JT distortions of 4.15 \% and 4.6 \%, respectively. We
note that GGA+DMFT calculations correctly describe both electronic and
structural properties of paramagnetic KCuF$_3$. This shows that the JT 
distortion in paramagnetic KCuF$_3$ is caused by electronic correlations.

\begin{figure}[tbp!]
\centerline{\includegraphics[width=0.45\textwidth,clip]{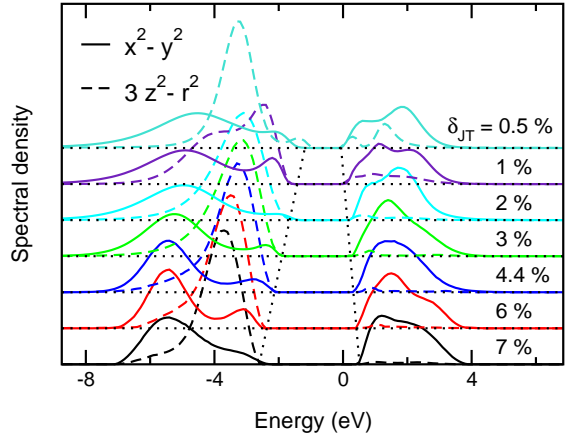}}
\caption{(color online) Orbitally resolved Cu $e_g$ spectral densities of
paramagnetic KCuF$_3$ as obtained by GGA+DMFT(QMC) for different values of
the JT distortion. The resulting orbitally resolved spectral density which 
is shown here by solid [dashed] line is predominantly of $x^2-y^2$ 
[$3z^2-r^2$] character (in the local frame \protect\cite{dif_orb}). }
\label{fig:kcuf_spectra}
\end{figure}

\begin{figure}[tbp!]
\includegraphics[width=0.3\textwidth]{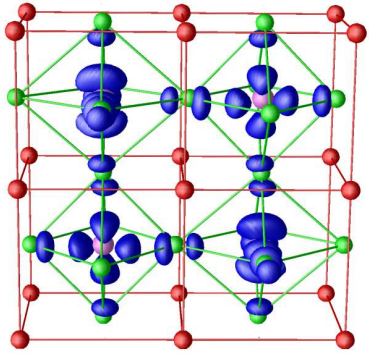}
\caption{(color online) $I4/mcm$ primitive cell and hole orbital order as
obtained by the GGA+DMFT calculation for paramagnetic KCuF$_3$ with 
$\protect\delta_{\mathrm{JT}}=4.4$ \%. The fluorine atoms and fluorine octahedra are 
shown in
green, the potassium in red, and the Wannier Cu $e_g$ charge density in
blue. The local coordinate system is chosen with the \emph{z} direction
defined along the longest Cu-F bond of the CuF$_6$ octahedron. }
\label{fig:kcuf_oo}
\end{figure}

Figure~\ref{fig:kcuf_spectra} shows the spectral density of paramagnetic 
KCuF$_3$, obtained from the QMC data by the maximum entropy method for 
several values of the JT distortion $\delta_{\mathrm{JT}}$. Most importantly, 
a paramagnetic
insulating state with a substantial orbital polarization is obtained for all
$\delta_{\mathrm{JT}}$ considered here. The energy gap is in the range 1.5--3.5 eV,
and increases with increasing of $\delta_{\mathrm{JT}}$. The sharp feature in the
spectral density at about $-3$ eV corresponds to the fully occupied $3z^2-r^2
$ orbital, \cite{dif_orb} whereas the lower and upper Hubbard bands are
predominantly of $x^2-y^2$ character and are located at $-5.5$ eV and 1.8
eV, respectively. The corresponding Cu $e_g$ Wannier charge density
calculated for the experimental value of JT distortion of 4.4 \% is
presented in Fig.~\ref{fig:kcuf_oo}. The GGA+DMFT results clearly show an
alternating occupation of the Cu $d_{x^2-z^2}$ and $d_{y^2-z^2}$ hole
orbitals, corresponding to the occupation of a $x^2-y^2$ hole orbital in the
local coordinate system, \cite{dif_orb} which implies antiferro-orbital
order.


The above calculations have been performed for the paramagnetic phase of 
KCuF$_3$. The N\' eel temperature ($T_N\sim22-38$ K for different types of
orbital order \cite{HS69}) is much lower than the temperature of present
calculations. However, it is known that the ordering temperature might be
overestimated by the single-site DMFT approximation, as is common for
mean-field theories. To prove the stability of the paramagnetic solution at high
temperatures (with respect to the $A$-type antiferromagnetic one) we have
carried out spin-polarized GGA+DMFT calculation at $T=560$ K. This
calculation has been performed for the $A$-type antiferromagnetic structure
using the experimental room-temperature crystal structure of KCuF$_3$ with 
$\delta_{\mathrm{JT}}=4.4$ \%. However, in agreement with experiment, the 
calculation gives paramagnetic insulating solution with the orbital order
as it has been found above. 

\subsection*{C.  Optimized structure}

To proceed further, we perform a structural optimization of paramagnetic 
KCuF$_3$. For simplicity, the optimization was performed only for two 
independent structural parameters, the lattice constant $a$ and the JT 
distortion $\delta_{\mathrm{JT}}$, keeping the space group symmetry 
(tetragonal $I4/mcm$) and the experimental value of the unit cell volume 
(taken at the ambient pressure at room temperature) unchanged. \cite{BM90}

\begin{figure}[htb]
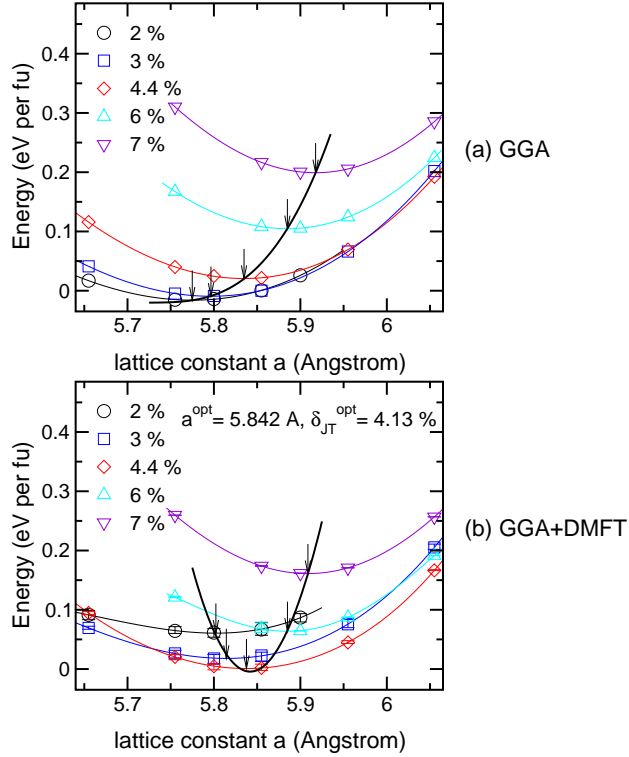

\centerline{\includegraphics[width=0.5\textwidth,clip]{etot_a_gga.eps}}
\centerline{\includegraphics[width=0.5\textwidth,clip]{etot_a_ggadmft.eps}}
\caption{(color online) Comparison of the total energies of paramagnetic 
KCuF$_3$ computed by GGA (a) and GGA+DMFT(QMC) (b) for different values of 
the JT distortion $\protect\delta_{\mathrm{JT}}$ as a function of the lattice 
constant $a$. Result of the total energy variation is marked by black arrows. 
Error bars indicate the statistical error of the DMFT(QMC) calculations. }
\label{fig:kcuf_opt_en}
\end{figure}

The calculations have been performed in two steps. In the first, we
calculate non-magnetic GGA electronic structure for different values 
of $\delta_{\mathrm{JT}}$ and lattice constant $a$. Note that in order 
to keep the unit cell volume constant, the value of tetragonal distortion 
($c/a$) was changed accordingly. In Figure~\ref{fig:kcuf_opt_en} (a) we plot 
the total energies obtained by GGA for different JT distortion 
$\delta_{\mathrm{JT}}$ as a function of
the lattice constant $a$. The data points were further interpolated by
smooth curvies, whereas the result of the total energy variation --- the 
line that connects the minima of the corresponding curves --- is marked 
by black arrows. We note that the result of the GGA structural optimization, 
the variation of the total energy, is seen to be constant for  
$\delta_{\mathrm{JT}}< 2$ \% with the end point at $a\sim 5.75$ \AA. 
This implies the absence of the cooperative JT distortion and results in 
a nearly cubic ($c/a \approx 1.0$) unit cell, which is in clear contradiction
to experiment. \cite{BM90}

In the second step, we construct the effective low-energy Hamiltonian for
the partially filled Cu $e_g$ orbitals for each value of the JT distortion $%
\delta_{\mathrm{JT}}$ and the lattice constant $a$ considered here, and compute the
corresponding total energies using GGA+DMFT approach 
(see Fig.~\ref{fig:kcuf_opt_en} (b)). \cite{LB08} In contrast to the structural 
optimization
within GGA, the inclusion of the electronic correlations among the partially
filled Cu $e_g$ states in the GGA+DMFT method not only correctly describes
the spectral properties, but also leads to a very prominent minimum in the
resulting total energy variation. The minimum is located at the value $%
a=5.842$ \AA\ and $\delta_{\mathrm{JT}}\approx 4.13$ \%, which is in 
excellent agreement with experimental value $a=5.855$ \AA\ and 
$\delta_{\mathrm{JT}} \approx 4.4$ \%. Note that in contrast to GGA, the 
structural optimization within GGA+DMFT also correctly predicts the 
tetragonal compression of the unit cell with $c/a \approx 0.95$. \cite{BM90}

\section*{IV.  APPLICATION TO LaMnO$_3$}

\subsection*{A.  Crystal structure and magnetic properties}

LaMnO$_3$ is another prototype of a material with a cooperative Jahn-Teller 
distortion and
orbital order. It stands in line with the colossal magnetoresistance
manganites whose parent compound it is. \cite{RevCMR} At ambient pressure and
temperature it
has an orthorhombic GdFeO$_3$-like crystal structure with space group $Pnma$
and four formula units (20 atoms) per primitive cell. \cite{EL71} The Mn ions
have octahedral oxygen surrounding and are in a high-spin $3d^4$ electronic
configuration due to Hund's rule coupling, with three electrons in the 
$t_{2g\uparrow}$ orbitals and a single electron in an $e_{g\uparrow}$ orbital
($t_{2g}^3 e_g^1$ orbital configuration). There are two types of structural
instabilities which give rise to the changes relative to the cubic
perovskite structure. The first is a JT instability due to the orbital
degeneracy that lifts the cubic degeneracy at Mn $e_g$ states due to
developing the cooperative JT distortion of the MnO$_6$ octahedra. The
second is related to a large ion-size misfit parameter 
$\sqrt{2} (R_O + R_{Mn})/(R_O + R_{La})$ 
which favors rotations of the octahedra to
accommodate a more efficient unit cell space filling. 
$R_{Mn}$, $R_{La}$, and $R_{O}$ are the ionic radii of Mn, La, and O 
ions, respectively. The
cooperative JT distortion lifts the $e_g$-orbital degeneracy and leads to an
alternating occupation of $d_{3x^2-r^2}$ and $d_{3y^2-r^2}$ electron
orbitals in the $ab$ plane (antiferro-orbital ordering) and to a tetragonal
compression of the unit cell. The rotations of the octahedra lower
the symmetry further, finally leading to the orthorhombic unit cell. 
In the paramagnetic $Pnma$ phase the JT distortion experimentally
persists up to $T_{JT} \approx 750$ K. At this temperature,
LaMnO$_3$ undergoes a structural phase transition, \cite{RH98} with volume
collapse \cite{CF03} to a nearly cubic structure in which orbital order and
JT distortion vanish. \cite{RH98} A quenching of the JT distortion has also
been reported in the room-temperature paramagnetic phase
under hydrostatic pressure $\sim 18$ GPa. At $\sim 32$ GPa it is followed 
by an insulator-metal transition. \cite{LA01}

At temperatures $T < T_N\sim 140$ K, which are much lower than $T_{JT}\sim 750$ K
where the JT distortion vanishes, LaMnO$_3$ shows $A$-type long-range
antiferromagnetic order consistent with the Goodenough-Kanamori-Anderson
rules for a superexchange interaction with $d_{3x^2-r^2}$/$d_{3y^2-r^2}$
antiferro-orbital order. \cite{EL71,RH98} This is also found in spin
polarized LDA/GGA and LDA+U calculations using the experimental
values of the crystal structure parameters. In this particular case, both
magnetic LDA/GGA and LDA+U calculations result in the qualitatively
correct insulating ground state with long-range $A$-type
antiferromagnetic and antiferro-orbital order. \cite{RK02,SM97} However,
the subsequent structural optimization within the magnetic LDA/GGA
calculations results in a \emph{metallic} solution with \emph{reduced} JT
distortion. \cite{SM97,TB05} In this situation, only the LDA+U scheme is
found to give, at equilibrium, the correct insulating character of the
low-temperature antiferromagnetic phase and a JT distortion in satisfactory
agreement with experiment. \cite{TB05} Nevertheless, we have to note again
that application of this approach is limited to temperatures below $T_N$.
Therefore, LDA+U cannot describe the properties of LaMnO$_3$ at $T>T_N$ and,
in particular, at room temperature, where LaMnO$_3$ is a correlated 
\emph{paramagnetic} insulator with a robust JT distortion. The electronic
properties of paramagnetic LaMnO$_3$ have already been studied
within the LDA+DMFT approach. In particular,
Pruschke and Z\" olfl \cite{PZ00} studied the electronic and magnetic
properties and found an additional increase of the orbital polarization
below $T_N$. Yamasaki \emph{et al.} \cite{YF06} examined the electronic
structure in order to address the origin of the high-pressure
metal-insulator transition. Pavarini and Koch \cite{PK09} investigated
the temperature dependence of the orbital polarization to find the origin of the
cooperative JT distortion and orbital order. However, no attempt has been
made to determine the structural properties and, in particular, the value of the
cooperative JT distortion of paramagnetic LaMnO$_3$ so far.

We present here the results of an application of the GGA+DMFT computational 
scheme
formulated in terms of plane-wave pseudopotentials \cite{LB08,TL08} to study
the electronic and structural properties of paramagnetic LaMnO$_3$. These are
the first results of a structural optimization where the 
stability of the cooperative JT distortion in paramagnetic 
LaMnO$_3$ was investigated. 
In principle, this application can be further extended to 
investigate the structural stability as a function of temperature.
Such a full structural
optimization will be interesting to study the disappearance of the JT 
distortion at $T \sim T_{JT}$. However, this is beyond the scope of the present
work.

\subsection*{B.  Electronic structure and orbital order}


In this section, we turn to the results of the GGA and GGA+DMFT electronic
structure calculations of paramagnetic LaMnO$_3$. In these calculations, we
have used the orthorhombic $Pnma$ crystal structure as reported by Elemans 
\emph{et al.}, with lattice constants $a=5.742$, $b=7.668$, and 
$c=5.532$ \AA. \cite{EL71} Similar to KCuF$_3$ we change the value of 
JT distortion $\delta_{\mathrm{JT}}$, which is now defined as the ratio 
between the difference of the long ($d_l$) and the short ($d_s$) bond 
distances and the mean Mn-O distance in the basal $ab$
plane, i.e., $\delta_{\mathrm{JT}}= 2(d_l-d_s)/(d_l+d_s)$. Structural
data \cite{EL71} yield $\delta_{\mathrm{JT}}= 0.138$. Note
that in the calculation we change only the parameter $\delta_{\mathrm{JT}}$ 
$(0 \leq \delta_{\mathrm{JT}} \leq 0.2)$ and keep the value of the 
MnO$_6$ octahedron tilting and rotation fixed.

Using the plane-wave pseudopotential calculation scheme, we calculated
the non-magnetic GGA electronic structure \cite{PB96,PSEUDO} of
paramagnetic LaMnO$_3$. We employed the Perdew-Burke-Ernzerhof
exchange-correlation functional together with Vanderbilt ultrasoft
pseudopotentials,
including a nonlinear core correction to the exchange-correlation potential.
All calculations were carried out in a 20-atom orthorhombic $Pnma$ unit cell.
We used a kinetic-energy cutoff of 45 Ry for the plane-wave expansion of the
electronic states. The integration in reciprocal space was performed using a
[10,10,10] Monkhorst-Pack $k$-point grid.

\begin{figure}[tbp!]
\centerline{\includegraphics[width=0.45\textwidth,clip]{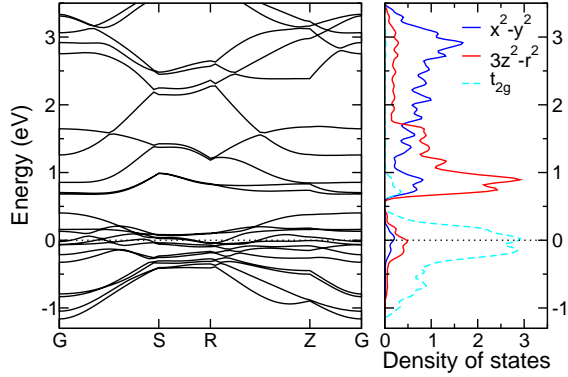}}

\caption{(color online) Orbitally resolved Mn 
$t_{2g}$ and $e_g$ spectral densities of paramagnetic LaMnO$_3$ as 
obtained by non-magnetic GGA for $\protect\delta_{\mathrm{JT}}= 0.138$. 
The zero of energy corresponds to the Fermi level.}
\label{fig:lamno_ggados}
\end{figure}

For all values of $\delta_{\mathrm{JT}}$ considered here the non-magnetic GGA
calculations give a metallic solution with a considerable orbital
polarization due to the crystal field splitting. Overall, these results
qualitatively agree with previous band structure calculations, \cite{YV06}
namely, that the GGA cannot describe a paramagnetic insulating behavior
which is found in experiment. We notice that even for the large 
$\delta_{\mathrm{JT}}$
value of 0.2 ($\sim 45$ \% larger than found in experiment \cite{EL71})
the GGA calculations predict a metal. The non-magnetic GGA density of states
and the corresponding band structure calculated for the JT distortion 
$\delta_{\mathrm{JT}}=0.138$ are presented in Fig.~\ref{fig:lamno_ggados}. 
%
%
In contrast to KCuF$_3$, the partially filled bands at the Fermi level now
originate from the Mn $t_{2g}$ orbitals. This is due to the $3d^4$
electronic configuration of Mn ions and the fictitious paramagnetic state
without local moments obtained by non-magnetic GGA. In fact, the Hund's 
rule coupling results in a strong on-site spin
polarization of the Mn $t_{2g}$ and $e_g$ orbitals. Therefore, the $t_{2g
\uparrow}$ states become completely occupied and located below the Fermi
level, while the remaining electron fills the $e_{g \uparrow}$ states. 
An increase of the JT distortion results in considerable enhancement of the
crystal field splitting between $x^2-y^2$ and $3z^2- r^2$ bands (in the
local frame \cite{dif_orb_lamno}) which reaches $\sim 1.1$ eV for 
$\delta_{\mathrm{JT}}=0.138$. The overall $e_g$ band width is about 
2.8-3.0 eV, which is
remarkably much smaller than the estimates of the Coulomb interaction parameter
U found in the literature. \cite{YF06,YV06}

\begin{figure}[tbp!]
\centerline{\includegraphics[width=0.45\textwidth,clip]{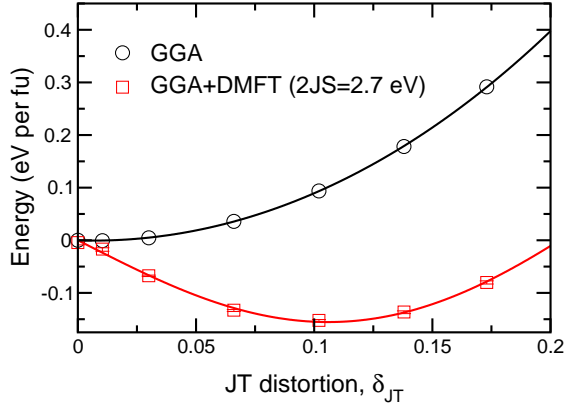}}
\caption{(color online) Comparison of the total energies of paramagnetic
LaMnO$_3$ computed by GGA and GGA+DMFT(QMC) as a function of the JT
distortion. Error bars indicate the statistical error of the DMFT(QMC)
calculations. }
\label{fig:lamno_energy}
\end{figure}

In Fig.~\ref{fig:lamno_energy} we display our results for the GGA total
energy as a function of the JT distortion $\delta_{\mathrm{JT}}$. In contrast to
experiment, the non-magnetic GGA calculations give a metallic solution
without cooperative JT distortion. Thus, the GGA total energy is almost
parabolic which implies the absence of a cooperative JT distortion and is in 
clear
contradiction to experiment. \cite{EL71,RH98} As in the case of KCuF$_3$ this
shows the importance of electronic correlations, without which 
the experimentally observed orbitally ordered, insulating state in paramagnetic
LaMnO$_3$ cannot be explained.

\begin{figure}[tbp!]
\includegraphics[width=0.22\textwidth]{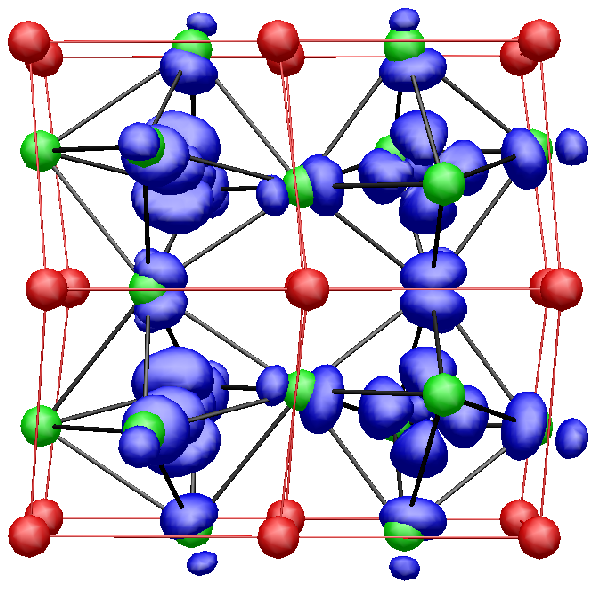} 
\includegraphics[width=0.22\textwidth]{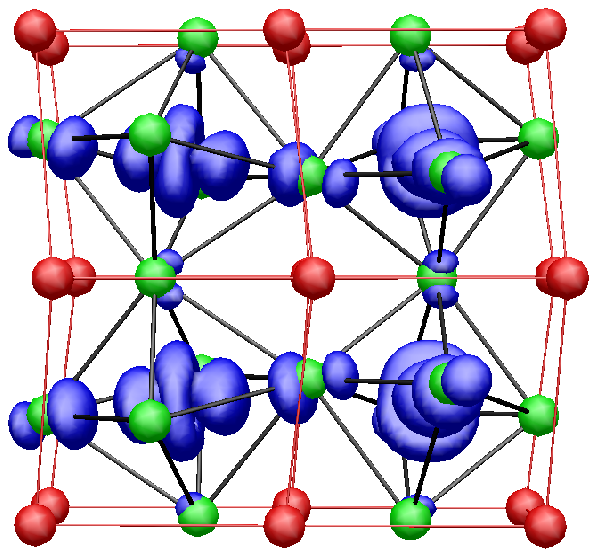} 
\caption{(color online) $Pnma$ primitive cell and the Wannier $e_g$ orbitals 
($x^2-y^2$ and $3z^2-r^2$ in left and right, respectively 
\protect\cite{dif_orb_lamno}) for LaMnO$_3$ with 
$\protect\delta_{\mathrm{JT}}= 0.138$ according to the non-magnetic GGA 
calculation. The oxygen atoms and oxygen octahedra are shown in green, the 
lanthanum in red, and the Wannier Mn $e_g$ charge density in blue. }
\label{fig:lamno_wan}
\end{figure}

Next, we turn to the GGA+DMFT results where we treat the Mn $e_g$ orbitals
as correlated orbitals. Using pseudopotential plane-wave approach, we
perform a projection onto atomic-centered symmetry-constrained Mn $e_g$
Wannier orbitals, \cite{TL08} which are shown in Fig.~\ref{fig:lamno_wan}. 
%
In this calculation we assume that three (among the $3d^4$ electronic
configuration) electrons are localized in the $t_{2g}$ orbitals. Therefore, 
they are 
treated as classical spins $S$, with a random orientation above $T_N$ (i.e., 
there is no
correlation between different Mn sites), which couple to the $e_g$
electron with an energy $JS$. This coupling can be estimated as the energy 
of the splitting
of the $e_{g \uparrow}$ and $e_{g \downarrow}$ bands in the ferromagnetic
band-structure calculations and gives an additional term in the Hamiltonian
(1), namely,
\begin{eqnarray}
{\hat H} & = & {\hat H_{DFT}} + U\sum_{im} \hat n_{im\uparrow} \hat
n_{im\downarrow} - JS \sum_{im} (\hat n_{im\uparrow} - \hat n_{im\downarrow}
)  \notag \\
& + & \sum_{i m\neq m^{\prime} \sigma \sigma^{\prime}} (V - \delta_{\sigma
\sigma^{\prime}} J) \hat n_{i m\sigma} \hat n_{i m^{\prime} \sigma^{\prime}}
- {\hat H_{DC}}.
\end{eqnarray}
This corresponds to the ferromagnetic Kondo-lattice model (KLM)
Hamiltonian with an on-site Coulomb repulsion between $e_g$ electrons,
which has been intensively studied as a possible microscopic model
to explain colossal magnetoresistance in manganites. \cite{MS96,HV00} We note
that in order to calculate the total energy one needs to modify 
 Eq.~2 by adding the expectation value of the $JS$ term which describes the
total energy gain due to the spin polarization of the $e_g$ orbitals at the
Mn site.

We take the local Coulomb repulsion $U=5$ eV, the Hund's rule exchange 
$J=0.75$ eV, and $2JS=2.7$ eV from the literature \cite{YF06} and further
solve the many-body Hamiltonian (5) for each value of $\delta_{\mathrm{JT}}$ 
using the single-site DMFT with Hirsch-Fey quantum Monte Carlo (QMC) calculations.
\cite{HF86,Off-diag-elements} The calculations were again 
performed at $T=1160$ K ($\beta=10$ eV$^{-1}$), using 40 imaginary-time slices.

In Fig.~\ref{fig:lamno_energy} we present the result of the paramagnetic
GGA+DMFT computation of the total energy, where it is compared with the
results of the
non-magnetic GGA calculation. In contrast to the GGA result, the correlated
electron problem solved by GGA+DMFT approach gives a substantial total
energy gain of $\sim$ 150 meV per formula unit. This implies that the
cooperative JT distortion indeed persists up to high temperatures ($T>1000$ K), 
while in GGA a JT distortion does not occur at all. Taking
into account that the calculations have been performed for the
low-temperature crystal structure of LaMnO$_3$ \cite{EL71} this estimate (150 meV) 
is in good agreement with $T_{JT} \sim 750$ K at which the JT distortion 
vanishes. \cite{RH98,CF03}
However, the structural change as a function of temperature in LaMnO$_3$, 
as well as the
disappearance of the orbital-order and JT distortion \cite{RH98,CF03} remains
an open problem. The minimum of the GGA+DMFT
total energy is located at the value $\delta_{\mathrm{JT}} \sim 0.11$, which is also in
good agreement with the experimental value of 0.138. \cite{EL71,RH98} We
note that GGA+DMFT calculations correctly describe both electronic and
structural properties of paramagnetic LaMnO$_3$. This shows that the 
JT distortion in paramagnetic LaMnO$_3$ is caused by electronic correlations.

\begin{figure}[tbp!]
\centerline{\includegraphics[width=0.45\textwidth,clip]{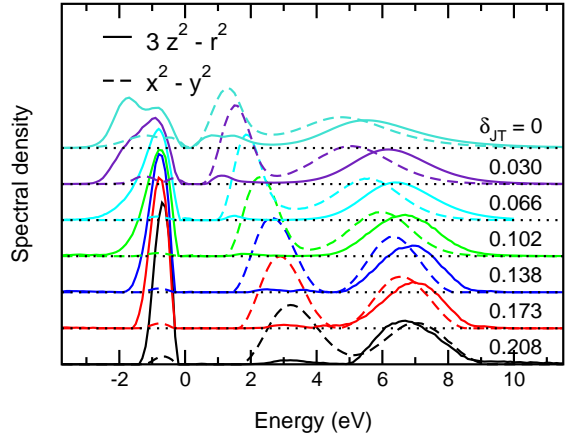}}
\caption{ (color online) Orbitally resolved Mn $e_g$ spectral densities of
paramagnetic LaMnO$_3$ as obtained by GGA+DMFT(QMC) for different values of
the JT distortion. The resulting orbitally resolved
spectral density shown by solid [dashed] line is predominantly of $3z^2-r^2$
[$x^2-y^2$] character (in the local frame \protect\cite{dif_orb_lamno}). }
\label{fig:lamno_spectra}
\end{figure}

\begin{figure}[tbp!]
\includegraphics[width=0.3\textwidth]{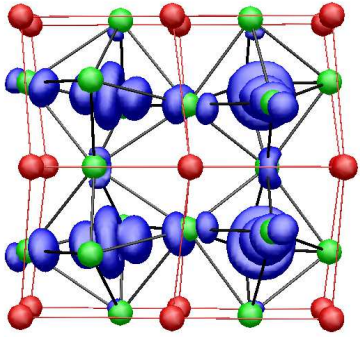}
\caption{(color online) $Pnma$ primitive cell and orbital order as obtained
by the GGA+DMFT calculation for paramagnetic LaMnO$_3$ with 
$\protect\delta_{\mathrm{JT}}=0.138$. The oxygen atoms and oxygen octahedra 
are shown in green, the lanthanum in red, and the Wannier Mn $e_g$ charge density 
in blue. 
}
\label{fig:lamno_oo}
\end{figure}

The spectral densities of paramagnetic LaMnO$_3$ calculated for several
values of the JT distortion $\delta_{\mathrm{JT}}$ using the maximum 
entropy analysis of the QMC data are shown in Fig.~\ref{fig:lamno_spectra}. 
For large $\delta_{\mathrm{JT}}$, we find a strong orbital polarization which 
gradually decreases
for decreasing JT distortion. The occupied part of the $e_g$ density
is located at about -2 -- -1 eV and corresponds to the $e_g$ states with
spin parallel to the $t_{2g}$ spin at that site. It has predominantly Mn 
$d_{3x^2-r^2}$ and $d_{3y^2-r^2}$ orbital character with a considerable 
admixture
of $d_{z^2-r^2}$ for small JT distortions. The energy gap is about 2 eV for
large $\delta_{\mathrm{JT}}$ and considerably decreases with decreasing 
$\delta_{\mathrm{JT}}$, resulting in a pseudogap behavior at the Fermi 
level for $\delta_{\mathrm{JT}}=0$. 
In Fig.~\ref{fig:lamno_oo} we show the corresponding Mn $e_g$ Wannier charge
density computed for the experimental JT distortion value of 
$\delta_{\mathrm{JT}}=0.138$. The result clearly show an alternating 
occupation of the
Mn $d_{3x^2-r^2}$ and $d_{3y^2-r^2}$ orbitals, corresponding to the
occupation of a $3z^2-r^2$ orbital in the local frame, \cite{dif_orb_lamno} 
which implies antiferro-orbital order. Thus, in agreement with experiment, 
the calculations give a paramagnetic insulating solution with antiferro-orbital
order and stable JT distortion.

\section*{V.  SUMMARY AND CONCLUSIONS}

In conclusion, by formulating GGA+DMFT --- the combination of the 
\emph{ab initio} band structure calculation technique GGA with the dynamical
mean-field theory --- in terms of plane-wave pseudopotentials, \cite{LB08,TL08}
we constructed a robust computational scheme for the investigation of
complex materials with strong electronic interactions.
Most importantly, the computational scheme presented here allows us to
explain correlation-induced structural transformations, shifts of equilibrium 
atomic positions and changes in the lattice structure, and to perform a 
structural optimization 
of \emph{paramagnetic} solids. We presented applications 
of this approach to two prototypical Jahn-Teller materials, KCuF$_3$
and LaMnO$_3$, and computed the orbital order and
cooperative JT distortion in these compounds. In particular, 
our results obtained for the paramagnetic phase of KCuF$_3$ and
LaMnO$_3$, namely an equilibrium Jahn-Teller distortion $\delta_{\mathrm{JT}}$ 
of 4.2 \% and 0.11,
respectively, and antiferro-orbital order, agree well with experiment. 
The present approach overcomes the limitations of the LDA+U method and is able 
to determine correlation-induced structural transformations in both 
paramagnetic and long-range magnetically ordered solids, and 
can thus be employed for the lattice optimization and molecular-dynamic simulations of
these systems. The GGA+DMFT scheme presented in this paper opens the way for
fully microscopic investigations of the structural properties of strongly
correlated electron materials such as lattice instabilities observed at
correlation-induced metal-insulator transitions.

\section*{ACKNOWLEDGMENTS} 

We thank M. Altarelli, J. Deisenhofer, D. Khomskii, K. Samwer, S. Streltsov,
N. Stoji\' c, and G. Trimarchi for valuable discussions. Support by the
Russian Foundation for Basic Research under Grant No. RFFI-07-02-00041, the
Deutsche Forschungsgemeinschaft through 
Sonderforschungsbereich 484, and the Light Source Theory
Network, LighTnet of the EU is gratefully acknowledged. Calculations were
performed at the CINECA supercomputing center in Bologna.


\begin{thebibliography}{99}

\bibitem{Rev} M. Imada, A. Fujimori, and Y. Tokura, Rev. Mod. Phys. \textbf{%
70}, 1039 (1998); Y. Tokura and N. Nagaosa, Science \textbf{288}, 462
(2000); E. Dagotto, Science \textbf{309}, 257 (2005).

\bibitem{LDA} R. O. Jones and O. Gunnarsson, Rev. Mod. Phys. \textbf{61},
689 (1989).

\bibitem{PB96} J. P. Perdew, K. Burke, and M. Ernzerhof, Phys. Rev. Lett.
\textbf{77}, 3865 (1996).

\bibitem{GGALDA} In general, GGA tends to give better results than LDA for
the electronic and structural properties of complex oxides and related
materials. See, D. R. Hamann, Phys. Rev. Lett. \textbf{76}, 660 (1996) and
H. Sawada, Y. Morikawa, K. Terakura, and N. Hamada, 
Phys. Rev. B \textbf{56}, 12154 (1997).

\bibitem{AZ91} V. I. Anisimov, J. Zaanen, and O. K. Andersen, Phys. Rev. B
\textbf{44}, 943 (1991); V. I. Anisimov, F. Aryasetiawan, and A. I.
Lichtenstein, J. Phys.: Condens. Matter \textbf{9}, 767 (1997).

\bibitem{LA95} A. I. Liechtenstein, V. I. Anisimov, and J. Zaanen, Phys.
Rev. B \textbf{52}, R5467 (1995).

\bibitem{RevFLAPW} H. J. Jansen and A. J. Freeman, Phys. Rev. B \textbf{30},
561 (1984).

\bibitem{RevPWSCF} S. Baroni, S. de Gironcoli, A. Dal Corso, and P. Giannozzi,
Rev. Mod. Phys. \textbf{73}, 515 (2001);
P. Giannozzi, S. Baroni, N. Bonini, M. Calandra, R. Car, C. Cavazzoni, 
D. Ceresoli, G. L. Chiarotti, M. Cococcioni, I. Dabo, A. Dal Corso, S. Fabris, 
G. Fratesi, S. de Gironcoli, R. Gebauer, U. Gerstmann, C. Gougoussis, A. Kokalj, 
M. Lazzeri, L. Martin-Samos, N. Marzari, F. Mauri, R. Mazzarello, S. Paolini, 
A. Pasquarello, L. Paulatto, C. Sbraccia, S. Scandolo, G. Sclauzero, A. P. Seitsonen, 
A. Smogunov, P. Umari, and R. M. Wentzcovitch, J. Phys. Condens. Matter \textbf{21}, 395502 (2009).

\bibitem{RevWIEN2k} P. Blaha, K. Schwarz, G. Madsen, D. Kvasnicka, and J.
Luitz, WIEN2k, \emph{An Augmented Plane Wave + Local Orbitals Program for
Calculating Crystal Properties} (Karlheinz Schwarz, Techn. Universit\" at
Wien, Austria), 2001.

\bibitem{RevVASP} G. Kresse and J. Hafner, Phys. Rev. B \textbf{47}, 558
(1993); G. Kresse and J. Hafner, Phys. Rev. B \textbf{49}, 14251 (1994); G. Kresse and 
J. Furthm\" uller, Phys. Rev. B \textbf{54}, 11169 (1996).

\bibitem{RevABINIT} X. Gonze, J.-M. Beuken, R. Caracas, F. Detraux, M.
Fuchs, G.-M. Rignanese, L. Sindic, M. Verstraete, G. Zerah, F. Jollet, M.
Torrent, A. Roy, M. Mikami, Ph. Ghosez, J.-Y. Raty, and D.C. Allan,
Computational Materials Science \textbf{25}, 478-492 (2002).

\bibitem{RevSIESTA} M. Soler, E. Artacho, J. D. Gale, A. 
Garc\'{\i}a, J. Junquera, P. Ordej\' on, and D. S\' anchez-Portal, J.
Phys.: Condens. Matter \textbf{14}, 2745 (2002).

\bibitem{DMFT} W. Metzner and D. Vollhardt, Phys. Rev. Lett. \textbf{62},
324 (1989); 
A. Georges, G. Kotliar, W. Krauth, and M. J. Rozenberg, Rev. Mod. Phys. \textbf{68}, 13
(1996); G. Kotliar and D. Vollhardt, Phys. Today \textbf{57}, No. 3, 53
(2004); G. Kotliar, S. Y. Savrasov, K. Haule, V. S. Oudovenko, O. Parcollet, and
C. A. Marianetti, Rev. Mod. Phys. \textbf{78}, 865 (2006).

\bibitem{DMFTmeth} V. I. Anisimov, A. I. Poteryaev, M. A. Korotin, 
A. O. Anokhin and G. Kotliar, J. Phys. Condens. Matt.
\textbf{9}, 7359 (1997); A. I. Lichtenstein and M. I. Katsnelson, Phys. Rev.
B \textbf{57}, 6884 (1998); 
K. Held, I. A. Nekrasov, G. Keller, V. Eyert, N. Bl\" umer, A. K. McMahan, R. T. Scalettar, 
Th. Pruschke, V. I. Anisimov, and D. Vollhardt, Psi-k Newsletter
\textbf{56}, 65 (2003); K. Held, I. A. Nekrasov, G. Keller, V. Eyert, 
N. Bl\" umer, A. K. McMahan, R. T. Scalettar, Th. Pruschke, V. I. Anisimov, and 
D. Vollhardt, Phys. Status Solidi B \textbf{243}, 2599 (2006).

\bibitem{DMFTcalc} K. Held, G. Keller, V. Eyert, D. Vollhardt, 
and V. I. Anisimov, Phys. Rev. Lett. \textbf{86}, 5345
(2001); E. Pavarini, S. Biermann, A. Poteryaev, A. I. Lichtenstein, 
A. Georges, and O. K. Andersen, Phys. Rev. Lett. \textbf{92}, 176403
(2004); A. I. Poteryaev, A. I. Lichtenstein, and G. Kotliar, Phys. Rev.
Lett. \textbf{93}, 086401 (2004); S. Biermann, A. Poteryaev, 
A. I. Lichtenstein, and A. Georges, Phys. Rev.
Lett. \textbf{94}, 026404 (2005); L. Chioncel, Ph. Mavropoulos, M. Lezai\' c, 
S. Bl\" ugel, E. Arrigoni, M. I. Katsnelson, and A. I. Lichtenstein, Phys. Rev.
Lett. \textbf{96}, 197203 (2006); J. Kunes, V. I. Anisimov, S. L. Skornyakov, 
A. V. Lukoyanov, and D. Vollhardt, Phys. Rev. Lett. \textbf{99}, 156404 (2007).

\bibitem{DMFTcalc+} M. I. Katsnelson and A. I. Lichtenstein Phys. Rev. B
\textbf{61}, 8906 (2000); A. I. Lichtenstein, M. I. Katsnelson, and G.
Kotliar, Phys. Rev. Lett. \textbf{87}, 067205 (2001); J. Braun, J. Min\' ar,
H. Ebert, M. I. Katsnelson, and A. I. Lichtenstein, Phys. Rev. Lett. \textbf{%
97}, 227601 (2006); A. Grechnev, I. Di Marco, M. I. Katsnelson, A. I.
Lichtenstein, J. Wills, and O. Eriksson, Phys. Rev. B \textbf{76}, 035107
(2007); S. Chadov, J. Min\'{a}r, M. I. Katsnelson, H. Ebert, D. K\"{o}%
dderitzsch, and A. I. Lichtenstein, Europhys. Lett. \textbf{82}, 37001
(2008).

\bibitem{LNMTO} O. K. Andersen, Phys. Rev. B \textbf{12}, 3060 (1975); O. K.
Andersen and T. Saha-Dasgupta, Phys. Rev. B \textbf{62}, R16219 (2000).

\bibitem{HM01} K. Held, A. K. McMahan, and R. T. Scalettar, Phys. Rev. Lett.
\textbf{87}, 276404 (2001); A. K. McMahan, K. Held, and R. T. Scalettar, 
Phys. Rev. B \textbf{67}, 075108 (2003).

\bibitem{AB06} B. Amadon, S. Biermann, A. Georges, and 
F. Aryasetiawan, Phys. Rev. Lett. \textbf{96}, 066402
(2006); L. V. Pourovskii, B. Amadon, S. Biermann, and A. Georges, Phys. Rev.
B \textbf{76}, 235101 (2007).

\bibitem{SKA01+} S. Y. Savrasov, G. Kotliar, and E. Abrahams, Nature
(London) \textbf{410}, 793 (2001); X. Dai, S. Y. Savrasov, G. Kotliar, 
A. Migliori, H. Ledbetter, and E. Abrahams, Science \textbf{300}, 
953 (2003); S. Y. Savrasov and G. Kotliar, Phys. Rev. B \textbf{69},
245101 (2004).

\bibitem{MnO} J. Kunes, A. V. Lukoyanov, V. I. Anisimov, R. T. Scalettar, 
and W. E. Pickett, Nature Materials 7, 198 (2008).

\bibitem{DM09+KS09}
The problem of equilibrium volume of number of simple elements has been
also recently addressed by I. Di Marco, J. Min\' ar, S. Chadov, M. I. Katsnelson, 
H. Ebert, and A. I. Lichtenstein [Phys. Rev. B \textbf{79}, 115111 (2009)] 
and A. Kutepov, S. Y. Savrasov, and G. Kotliar 
[Phys. Rev. B \textbf{80}, 041103 (R) (2009)].

\bibitem{JT37} H. A. Jahn and E. Teller, Proc. R. Soc. London Ser. A \textbf{%
161}, 220 (1937).

\bibitem{KK} D. I. Khomskii and K. I. Kugel, Solid State Comm. \textbf{13},
763 (1973); K. I. Kugel and D. I. Khomskii, Sov. Phys. Solid State \textbf{17%
}, 285 (1975); K. I. Kugel and D. I. Khomskii, Sov. Phys. JETP \textbf{52},
501 (1981); K. I. Kugel and D. I. Khomskii, Sov. Phys. Usp. \textbf{25}(4),
231 (1982).

\bibitem{LB08} I. Leonov, N. Binggeli, Dm. Korotin, V. I. Anisimov, N.
Stoji\' c, and D. Vollhardt, Phys. Rev. Lett. \textbf{101}, 096405 (2008).

\bibitem{TL08} G. Trimarchi, I. Leonov, N. Binggeli, Dm. Korotin, 
and V. I. Anisimov, J. Phys.: Condens. Matter \textbf{20}, 135227 (2008).

\bibitem{DK08} Dm. Korotin, A. V. Kozhevnikov, S. L. Skornyakov, 
I. Leonov, N. Binggeli, V. I. Anisimov, and G. Trimarchi, 
The European Physical Journal B \textbf{65}, 91 (2008).

\bibitem{AL08} B. Amadon, F. Lechermann, A. Georges, F. Jollet, T. O. Wehling, 
and A. I. Lichtenstein, Phys. Rev. B \textbf{77}, 205112 (2008).

\bibitem{LG06} For a formulation of LDA+DMFT within a mixed-basis
pseudopotential approach see F. Lechermann, A. Georges, A. Poteryaev, 
S. Biermann, M. Posternak, A. Yamasaki, and O. K. Andersen, Phys. Rev. B
\textbf{74}, 125120 (2006).

\bibitem{BM90} R. H. Buttner, E. N. Maslen, and N. Spadaccini, Acta Cryst. B
\textbf{46}, 131 (1990).

\bibitem{EL71} J. B. A. A. Elemans and B. van Laar, K. R. van der Veen,
and B. O. Loopstra, J. Phys. Chem. Solids \textbf{3}, 238 (1971).

\bibitem{RH98} J. Rodriguez-Carvajal, M. Hennion, F. Moussa, A. H. Moudden,
L. Pinsard, and A. Revcolevschi, Phys. Rev. B. \textbf{57}, R3189 (1998).

\bibitem{CF03} T. Chatterji, F. Fauth, B. Ouladdiaf, P. Mandal, and B.
Ghosh, Phys. Rev. B \textbf{68}, 052406 (2003).

\bibitem{MS96} A. J. Millis, B. I. Shraiman, and R. Mueller, Phys. Rev.
Lett. \textbf{77}, 175 (1996).

\bibitem{HV00} K. Held and D. Vollhardt, Phys. Rev. Lett. \textbf{84}, 5168
(2000).

\bibitem{AK05} V. I. Anisimov, D. E. Kondakov, A. V. Kozhevnikov, I. A. Nekrasov, 
Z. V. Pchelkina, J. W. Allen, S.-K. Mo, H.-D. Kim, P. Metcalf, S. Suga, A. Sekiyama, 
G. Keller, I. Leonov, X. Ren, and D. Vollhardt, Phys. Rev. B \textbf{71}, 125119 (2005).

\bibitem{MV97} N. Marzari and D. Vanderbilt, Phys. Rev. B \textbf{56}, 12847
(1997).

\bibitem{HF86} J. E. Hirsch and R. M. Fye, Phys. Rev. Lett \textbf{56}, 2521
(1986).

\bibitem{KY67} S. Kadota, I. Yamada, S. Yoneyama, and K. Hirakawa, J. Phys.
Soc. Jpn. 23, 751 (1967).

\bibitem{O69} A. Okazaki, J. Phys. Soc. Jpn. \textbf{26}, 870 (1969);
A. Okazaki, J. Phys. Soc. Jpn. \textbf{27}, 518B (1969).

\bibitem{G63} J. B. Goodenough, \textit{Magnetism and the Chemical Bond}
(Interscience, New York, 1963).

\bibitem{MK02} J. E. Medvedeva, M. A. Korotin, V. I. Anisimov, 
and A. J. Freeman, Phys. Rev. B \textbf{65}, 172413 (2002).

\bibitem{PK08} E. Pavarini, E. Koch, and A. I. Lichtenstein, Phys. Rev.
Lett. \textbf{101}, 266405 (2008).

\bibitem{BA04} N. Binggeli and M. Altarelli, Phys. Rev. B \textbf{70},
085117 (2004).

\bibitem{HS69} M. T. Hutchings, E. J. Samuelsen, G. Shirane, and 
K. Hirakawa, Phys. Rev. \textbf{188}, 919 (1969).

\bibitem{U91} T. Ueda, K. Sugawara, T. Kondo, and I. Yamada, 
Solid State Commun. \textbf{80}, 801 (1991).

\bibitem{Y89} I. Yamada, H. Fujii, and M. Hidaka, J. Phys. Condens. 
Matter \textbf{1}, 3397 (1989).

\bibitem{EZ08} M. V. Eremin, D. V. Zakharov, H.-A. Krug von Nidda, R. M.
Eremina, A. Shuvaev, A. Pimenov, P. Ghigna, J. Deisenhofer, and A. Loidl,
Phys. Rev. Lett. \textbf{101}, 147601 (2008); J. Deisenhofer, I. Leonov, M.
V. Eremin, Ch. Kant, P. Ghigna, F. Mayr, V. V. Iglamov, V. I. Anisimov, and
D. van der Marel, \emph{ibid.} \textbf{101}, 157406 (2008).

\bibitem{PC02} L. Paolasini, R. Caciuffo, A. Sollier, P. Ghigna, and M.
Altarelli, Phys. Rev. Lett. \textbf{88}, 106403 (2002); R. Caciuffo, L.
Paolasini, A. Sollier, P. Ghigna, E. Pavarini, J. van den Brink, and M.
Altarelli, Phys. Rev. B \textbf{65}, 174425 (2002).

\bibitem{PSEUDO} Calculations have been performed using the Quantum 
ESPRESSO package, see Ref.~\onlinecite{RevPWSCF}, 
URL http://www.quantum-espresso.org.


\bibitem{dif_orb} The local coordinate system is chosen with the \emph{z} 
direction defined along the longest (in $ab$ plane) Cu-F bond of the 
CuF$_6$ octahedron.

\bibitem{Off-diag-elements} To simplify the computation we neglected the
orbital off-diagonal elements of the local Green's function by applying an
additional transformation into the local basis set with a diagonal density
matrix during each DMFT iteration.

\bibitem{RevCMR} For a review, see, for example, Y. Tokura, Ed., \emph{Colossal
Magnetoresistive Oxides} (Gordon and Breach Science, New York, 2000), and
references therein.

\bibitem{LA01} I. Loa, P. Adler, A. Grzechnik, K. Syassen, U. Schwarz, M..
Hanfland, G. Kh. Rozenberg, P. Gorodetsky, and M. P. Pasternak, Phys. Rev.
Lett. \textbf{87}, 125501 (2001).

\bibitem{RK02} P. Ravindran, A. Kjekshus, H. Fjellvag, A. Delin, and O.
Eriksson, Phys. Rev. B \textbf{65}, 064445 (2002); and references therein.

\bibitem{SM97} H. Sawada, Y. Morikawa, K. Terakura, and N. Hamada, Phys.
Rev. B \textbf{56}, 12154 (1997).

\bibitem{TB05} G. Trimarchi and N. Binggeli, Phys. Rev. B \textbf{71},
035101 (2005).

\bibitem{PZ00} Th. Pruschke and M.B. Z\" olfl, \emph{Advances in Solid State Physics} 
\textbf{40}, 251 (2000);
See also R. Peters and Th. Pruschke, cond-mat/0908.3990, where the authors 
discuss the interplay of orbital and spin degrees of 
freedom in the two orbital Hubbard model near quarter filling.

\bibitem{YF06} A. Yamasaki, M. Feldbacher, Y.-F. Yang, O. K. Andersen, and
K. Held, Phys. Rev. Lett. \textbf{96}, 166401 (2006); K. Held, O. K.
Andersen, M. Feldbacher, A. Yamasaki, and Y.-F. Yang, J. Phys.: Condens.
Matter \textbf{20}, 064202 (2008).

\bibitem{PK09} E. Pavarini and E. Koch, cond-mat/0904.4603.

\bibitem{YV06} W.-G. Yin, D. Volja, and W. Ku, Phys. Rev. Lett. \textbf{96},
116405 (2006).

\bibitem{dif_orb_lamno} The local coordinate system is chosen such that
the GGA Mn $3d$ density matrix has a diagonal form.


\end{thebibliography}
\end{document}